\documentclass[apj]{emulateapj}
\usepackage{epsfig}
\usepackage{amsmath}
\usepackage{natbib}
\usepackage{graphicx,subfigure}
\usepackage{calc}
\usepackage{verbatim}
\usepackage{tabularx}
\usepackage{xcolor}

\newcommand{\ztwo}{$z\sim2 \hspace{4pt}$}

\newcommand{\ls}{\hspace{2pt}}
\newcommand{\ha}{H$\alpha$\ls}				    
\newcommand{\nii}{[N{\small II}]\ls}                                       
\newcommand{\sii}{[S{\small II}]\ls}                                       
\newcommand{\oii}{[O{\small II}]}

\newcommand{\niiha}{[N{\small II}]/\ha}				    

\newcommand{\msun}{M$_{\odot}$} 			   
\newcommand{\sigsfrunits}{M$_{\odot} yr^{-1} kpc^{-2}$}
\newcommand{\mstar}{M$_{*}$} 				    
\newcommand{\mgas}{M$_{gas}$}

\newcommand{\sigsfr}{$\Sigma_{SFR}$}
\newcommand{\rhalf}{R$_{1/2}$}

\newcommand{\vrot}{v$_{rot}$}
\newcommand{\sn}{$\sigma_{0}$}

\newcommand{\kms}{km~s$^{-1}$}			              

\shorttitle{Dispersion dominated \ztwo galaxies}
\shortauthors{S. F. Newman et al.}

\begin{document}

\title{The SINS/zC-SINF survey of \ztwo galaxy kinematics: the nature of dispersion dominated galaxies$^{*}$}
\author{Sarah F. Newman\footnotemark[1,17], Reinhard Genzel\footnotemark[1,2,3], Natascha M. F\"orster Schreiber\footnotemark[2], Kristen Shapiro Griffin\footnotemark[4], Chiara Mancini\footnotemark[5],  Simon J. Lilly\footnotemark[6], Alvio Renzini\footnotemark[5], Nicolas Bouch\'e\footnotemark[7,8], Andreas Burkert\footnotemark[9], Peter Buschkamp\footnotemark[2], C. Marcella Carollo\footnotemark[6], Giovanni Cresci\footnotemark[10], Ric Davies\footnotemark[2], Frank Eisenhauer\footnotemark[2], Shy Genel\footnotemark[11], Erin K. S. Hicks\footnotemark[12], Jaron Kurk\footnotemark[2], Dieter Lutz\footnotemark[2], Thorsten Naab\footnotemark[13], Yingjie Peng\footnotemark[6], Amiel Sternberg\footnotemark[14], Linda J. Tacconi\footnotemark[2], Stijn Wuyts\footnotemark[2], Gianni Zamorani\footnotemark[15], and Daniela Vergani\footnotemark[16]}

\footnotetext[*]{Based on observations at the Very Large Telescope (VLT) of the European Southern Observatory (ESO), Paranal, Chile (ESO program IDs 076.A-0527, 079.A-0341, 080.A-0330, 080.A-0339, 080.A-0635, 183.A-0781).}
\footnotetext[1]{Department of Astronomy, Campbell Hall, University of California, Berkeley, CA 94720, USA}
\footnotetext[2]{Max-Planck-Institut f\"ur extraterrestrische Physik (MPE), Giessenbachstr.1, D-85748 Garching, Germany}
\footnotetext[3]{Department of Physics, Le Conte Hall, University of California, Berkeley, CA 94720, USA}
\footnotetext[4]{Space Sciences Research Group, Northrop Grumman Aerospace Systems, Redondo Beach, CA 90278, USA}
\footnotetext[5]{Osservatorio Astronomico di Padova, Vicolo dell'Osservatorio 5, Padova, I-35122, Italy}
\footnotetext[6]{Institute of Astronomy, Department of Physics, Eidgen\"ossische Technische Hochschule, ETH Z\"urich, CH-8093, Switzerland}
\footnotetext[7]{Universit\'e de Toulouse; UPS-OMP; IRAP; Toulouse, France}
\footnotetext[8]{CNRS; IRAP; 14, avenue Edouard Belin, F-31400 Toulouse, France}
\footnotetext[9]{Universit\"ats-Sternwarte Ludwig-Maximilians-Universit\"at (USM), Scheinerstr. 1, M\"unchen, D-81679, Germany}
\footnotetext[10]{Istituto Nazionale di AstrofisicaÐOsservatorio Astronomico di Arcetri, Largo Enrico Fermi 5, I Ð 50125 Firenze, Italy}
\footnotetext[11]{Harvard-Smithsonian Center for Astrophysics, 60 Garden Street, Cambridge, MA 02138 USA}
\footnotetext[12]{Department of Astronomy, University of Washington, Box 351580, U.W., Seattle, WA 98195-1580, USA}
\footnotetext[13]{Max-Planck Institute for Astrophysics, Karl Schwarzschildstrasse 1, D-85748 Garching, Germany}
\footnotetext[14]{School of Physics and Astronomy, Tel Aviv University, Tel Aviv 69978, Israel}
\footnotetext[15]{INAF Osservatorio Astronomico di Bologna, Via Ranzani 1, 40127 Bologna, Italy}
\footnotetext[16]{INAF Istituto di Astrofisica Spaziale e Fisica Cosmica di Bologna, Via P. Gobetti 101, 40129 Bologna, Italy}
\footnotetext[17]{email: sfnewman@berkeley.edu}


\begin{abstract} 
We analyze the spectra, spatial distributions and kinematics of \ha, \nii and \sii \ls emission in a sample of 38, z $\sim$ 2.2 UV/optically selected star forming galaxies (SFGs) from the SINS \& zC-SINF surveys, 34 of which were observed in the adaptive optics mode of SINFONI and 30 of those contain data presented for the first time here. This is supplemented by kinematic data from 43 z $\sim$ 1--2.5 galaxies from the literature. None of these 81 galaxies is an obvious major merger. We find that the kinematic classification of high-z SFGs as `dispersion dominated' or `rotation dominated' correlates most strongly with their intrinsic sizes. Smaller galaxies are more likely `dispersion-dominated' for two main reasons: 1) The rotation velocity scales linearly with galaxy size but intrinsic velocity dispersion does not depend on size or may even increase in smaller galaxies, and as such, their ratio is systematically lower for smaller galaxies, and 2) Beam smearing strongly decreases large-scale velocity gradients and increases observed dispersion much more for galaxies with sizes at or below the resolution. Dispersion dominated SFGs may thus have intrinsic properties similar to `rotation dominated' SFGs, but are primarily more compact, lower mass, less metal enriched and may have higher gas fractions, plausibly because they represent an earlier evolutionary state.
\end{abstract}


\section{Introduction} 

At the peak of the galactic formation epoch at z$\sim$1--3, the rest-frame UV and \ha morphologies of most star forming galaxies near the `main sequence' of the stellar mass-star formation plane \citep[henceforth `SFGs':][]{Noe+07,Dad+07,Rod+11,Pen+10} are typically irregular and in some cases dominated by several giant (kpc-size) star forming clumps \citep{CowHuSon95,vdBer+96,ElmElmShe04,Elm+09,ElmElm05,ElmElm06,Gen+06,Gen+08,Gen+11,Law+07,Law+09,Law+12a,For+09,For+11a,Wuy+12,Swi+12a,Swi+12b}. The ionized gas kinematics of these clumpy galaxies range from rotationally supported disks, especially among the more massive (M$_{*} \geq$ a few 10$^{10}$ \msun) and bright (Ks AB $\leq$ 21.8) SFGs, to galaxies dominated by apparently random motions \citep{For+06,For+09,Law+07,Law+09,Gen+08,Gen+11,Wri+07,Wri+09,Sha+08,Cre+09,vSta+08,Epi+09,LemLam10,Jon+12,Wis+12,Swi+12a,Swi+12b,Epi+12}. The latter class, which are especially common among the smaller and lower mass systems, often have average intrinsic velocity dispersions (corrected for instrumental broadening and beam smearing, and determined here from the outer regions of the galaxies), \sn, larger than the inclination-corrected rotation velocities, \vrot \ls (hereafter, `dispersion dominated' SFGs) \citep{For+09}. Despite these irregular morphologies, the fraction of major mergers is probably less than 30\% \citep{Sha+08,For+09,Lop+12} and stellar mass maps of z$\sim$1--2 SFGs are typically smoother and more symmetric than the light distributions \citep{Wuy+12}. As a rule, rest-frame UV and optically-selected high-z SFGs exhibit large local random motions, with ratios of \vrot \ls to \sn \ls of less than 8. Hence even the rotationally dominated systems are thick \citep[H$_{z}\sim$ \ls 1 kpc:][]{ElmElm06} and highly turbulent. Observations of CO rotational line emission indicate that z$\sim$1--3 massive SFGs have large ($\sim$30--60\%) baryonic cold gas fractions and this cold molecular gas has a large velocity dispersion comparable to that of the warm ionized gas traced by \ha \citep{Dad+08,Dad+10b,Tac+08,Tac+10,Tac+12,Swi+11}. 

These basic observational properties (irregular morphologies, high gas fractions and large velocity dispersions) can be understood in a simple physical framework, in which global gravitational instability and fragmentation in marginally stable (Q$_{Toomre}\leq$ 1), gas-rich disks naturally leads to large turbulence and the formation of giant star forming clumps \citep{Nog99,Imm+04a,Imm+04b,BouElmElm07,ElmBouElm08a,Gen+08,Gen+11,DekSarCev09,Bou+10,GenDekCac12}. The buildup of these marginally stable disks in z\textgreater1 SFGs is plausibly fueled mainly by smooth accretion of gas and/or minor mergers \citep{Ker+05,Ker+09,DekBir06,Bow+06,KitWhi07,OcvPicTey08,Dave+08,DekSarCev09,Ose+10,CacDekGen12,Cev+12}.
 
While the buildup of early star forming disks may be explained by this scenario, it is less clear how the dispersion dominated galaxies fit in. There are several possible explanations for what they are \citep[see e.g.:][]{For+09,Law+07,Law+09,Law+12a,Jon+10,Epi+12}, including: 
\begin{itemize}

\item[] (1) an earlier evolutionary stage with higher gas fractions and lower masses, in which case the simple fragmentation scenario discussed above would lead to larger velocity dispersions,
\item[] (2) intrinsically smaller rotationally supported disks that masquerade as dispersion dominated because the instrumental `beam smearing' hides the rotational signal, 
\item[] (3) the result of dissipative major mergers that would drive chaotic motions, 
\item[] (4) giant clumps in face-on star forming disks, where the rest of the disk material has too low surface brightness to be detected, or
\item[] (5) some combination thereof.
\end{itemize}

In this paper we present and analyze new high-quality SINFONI/VLT integral field (IFU) spectroscopy with natural and laser guide star adaptive optics (AO) \citep{Eis+03,Bon+04} of 30 z$\sim$2 SFGs. We combine this with 4 AO and 4 seeing-limited data sets which have been presented previously by our group \citep{For+09,Cre+09,Gen+11}, but contain new analysis here. For these 38 SINS/zc-SINF galaxies, we find that 43\% were classified as `dispersion dominated' based on previous seeing limited observations and according to the criteria of \cite{For+09}. We add to our measurements another 35 AO data sets and 8 spatially well-resolved (with \rhalf \textgreater 4.5 kpc) seeing limited data sets of z = 1--2.5 SFGs from the literature. The combined data of 81 SFGs provide interesting new constraints on the nature of the dispersion dominated SFG population. We adopt a $\Lambda$CDM cosmology with $\Omega_{m}$=0.27, $\Omega_{b}$=0.046 and H$_{0}$=70 km/s/Mpc \citep{Kom+11}, as well as a \cite{Cha03} initial stellar mass function (IMF).
 
\section{Observations and Analysis}

\subsection{Source selection, observations and data reduction of SINS/zC-SINF galaxies}

Our base sample are 34 SFGs from the SINS and zC-SINF surveys of \ha+\nii \ls integral field spectroscopy in z $\sim$ 1.5--2.5 SFGs obtained with SINFONI on the ESO VLT that were observed in both seeing-limited and AO mode. These galaxies are drawn from the parent seeing-limited SINFONI samples described in \cite{For+09} and \cite{Man+11}. The main criteria for selecting these galaxies from the seeing-limited sample were a clear detection of \ha \ls in one hour seeing-limited SINFONI data, the proximity of a suitably bright AO reference star, and the goal of ultimately covering as much of the range in stellar mass, SFR and kinematic properties of the parent no-AO SINS/zC-SINF sample as possible. The resulting AO sample has nearly identical distributions in \mstar \ls and SFR as the parent no-AO sample in terms of range, median and offset from the main sequence \citep{For+13}. In addition to the SINS/zC-SINF AO sample, we include 4 large galaxies (\rhalf \textgreater 4.5 kpc) observed only in seeing-limited mode with sufficient signal to noise ratio (S/N) and spatial resolution to robustly determine the kinematics \citep{For+09,Man+11}. Adding these larger objects does not bias our results as we do not aim for a `complete' sample, merely one that has as large a range in galaxy properties and kinematic types as possible. Thus these galaxies only give us more information on rotationally-supported objects. We exclude obvious major mergers from our sample, but keep possible minor mergers, assuming that the central object in a minor merger retains its kinematic integrity to a significant extent. Note, however, that this criterion only eliminated 5 out of originally 86 objects (when including the additional datasets from the literature - see next Section).

The SINS and zC-SINF surveys were selected either from their U$_{n}$GR colors satisfying the `BX' criteria \citep{Ste+04,Ade+04,Erb+06,Law+09} or based on \textit{K} band imaging via the `BzK' criterion for 1.4 \textless z \textless 2.5 SFGs \citep{Dad+04}. In addition, a few galaxies were included based on their stellar masses and SFRs from the GMASS Spitzer 4$\mu$m survey \citep{Kur+09,Kur+13}, and one galaxy was selected from the GDDS survey based on a secure redshift and evidence for on-going star-formation from the UV data \citep{Abr+04}.

These galaxies sample the \ztwo SFG `main sequence' in the stellar mass -- star-formation rate plane between stellar masses of 10$^{9.2}$ and 10$^{11.5}$ \msun, and star formation rates between $\sim$13 and 850 \msun yr$^{-1}$ (see Figure 1), covering the same range in \mstar \ls and SFR as the parent samples. For the AO data, we employed SINFONI in the 0.05''x0.1'' pixel scale, with either laser guide star (LGS) or natural guide star (NGS) AO, resulting in angular resolutions of $\sim$0.20'' (equivalent to 1.7 kpc at \ztwo) full width at half maximum (FWHM) after median filtering by 3x3 pixels. The 4 larger rotationally supported systems were observed in SINFONI's seeing limited mode and 0.125''x0.25'' pixel scale, resulting in a median FWHM $\sim$0.56'', sufficient to resolve these SFGs \citep{For+09}. These sources are included in the analysis (and not other seeing-limited only galaxies), because their large spatial extents allow well-resolved kinematical analysis. The on-source integration times for each galaxy from our sample range between 2 and 23 hours, with a median of about 5.8 hours, resulting in high quality, spatially resolved spectra for most sources. This sample includes 6 AGN, identified by the presence of AGN signatures in their rest-UV spectrum, \niiha \ls ratio, and/or X-ray or MIPS 24$\mu m$ data when available \citep[e.g., see][]{For+09}. These would not affect our \sn \ls measurements, which are taken away from the center, but could potentially affect $\sigma_{tot}$ (the total Gaussian line width integrated over the source). A discussion of the properties of 9 individual galaxies of the AO sample can be found in several of our earlier SINS papers \citep{Gen+06,Gen+11,Cre+09,For+09,New+12a}, with new data here for 5 of those objects. A detailed discussion of the full AO sample will be presented in \cite{For+13}.
 
We used the software package SPRED and custom routines for optimizing the background/OH airglow subtraction for the data reduction. The point spread function (PSF) FWHM was measured by fitting a 2D Gaussian profile to the combined images of the PSF calibration star taken throughout the observations of a galaxy. More information on the specifics of these observations and the data reduction can be found in \cite{For+09,Man+11,For+13}.

We created \ha emission line, velocity and velocity dispersion maps from the reduced data cubes by using the Gaussian-fitting procedure LINEFIT \citep{Dav+11}, with errors derived from the noise cube. For more information on our standard SINS data reduction methods and analysis tools we refer to \cite{Sch+04,Dav+07,For+09,Man+11}. 

\subsection{Additional datasets from the literature}

We also include in our analysis 43 additional z $\sim$ 1--2.5 SFGs from the literature (Table 1) deriving from IFU datasets with beam-smearing corrected kinematic information. These include:

\begin{itemize}
\item 9 galaxies from the BX- (rest-UV color/magnitude) selected z = 1.5--2.5 AO sample from \cite{Law+09,Law+12c} taken with OSIRIS.
\item 6 galaxies from the BM-selected z = 1.5--1.7 AO sample from \cite{Wri+07,Wri+09} taken with OSIRIS.
\item 12 galaxies from the z = 1.3--1.45 AO sample from \cite{Wis+11} selected for strong \oii \ls emission from the Wiggle-Z Dark Energy (UV-selected) survey, also taken with OSIRIS.
\item 5 z $\sim$ 1.45 galaxies from \cite{Swi+12a,Swi+12b}, taken from the HiZELS imaging survey of the COSMOS and UDS fields, selected for narrow band \ha \ls flux and observed with SINFONI in the AO mode.
\item 8 z = 1-1.6 MASSIV galaxies selected from the VVDS spectroscopic survey observed in seeing-limited mode with SINFONI \citep{LemLam10,Epi+12, Con+12, Ver+12}. Here we only chose galaxies with \rhalf \textgreater 4.5 kpc, such that their kinematics could be well resolved without AO.
\item 3 additional z = 1.2--1.4 galaxies from \cite{Epi+12} observed in the AO mode of SINFONI.

\end{itemize}

\noindent These galaxies were selected such that they were in the redshift interval 1--2.5, were not classified as major mergers, and had either AO data or seeing-limited data with \rhalf \textgreater 4.5 kpc. We do not include z\textless1 studies \citep[e.g.]{Pue+07,Pue+09,Nei+08}, or z$\sim$1 slit spectroscopy \citep{Wei+06,Kas+07} in the analysis, but emphasize that these investigations also find evidence for large random motions.

\begin{table*}
\caption{Source Selection}
\scriptsize{
\begin{tabular*}{1\textwidth}{l c c c c l}
\\
\hline
\hline 
\textbf{Dataset} & \textbf{z} &  \textbf{Galaxies} & \textbf{AO?} & \textbf{Instrument} & \textbf{Reference} \vspace{2pt} \\
\hline 
SINS/zC-SINF & 1.5---2.5 & 34 & AO & SINFONI/VLT & this paper$^{1}$ \\
SINS/zC-SINF & 1.5---2.5 & 4 & seeing-lim. & SINFONI/VLT & \cite{For+09}$^{2}$ \\
BX/BM-selected & 1.5--2.5 & 15 & AO & OSIRIS/Keck & \cite{Wri+07,Wri+09,Law+09,Law+12c}$^{3}$ \\
Wiggle-Z & 1.3--1.45 & 12 & AO & OSIRIS/Keck & \cite{Wis+11} \\
HiZELS & 1.4--1.5 & 5 & AO & SINFONI/VLT & \cite{Swi+12a,Swi+12b} \\
MASSIV & 1.2--1.4 & 3 & AO & SINFONI/VLT & \cite{Epi+12, Con+12}$^{4}$ \\
MASSIV & 1--1.6 & 8 & seeing-lim. & SINFONI/VLT & \cite{LemLam10} and \\
& & & & & \cite{Epi+12,Con+12} \\

\hline

\end{tabular*}}
$^{1}$ Seeing-limited data of all of these galaxies has been presented in \cite{For+09} and \cite{Man+11}. Data of 9 of these galaxies have been presented in \cite{Gen+06,Gen+11,Cre+09,For+09,New+12a}, although 5 of these now contain new observations. The full AO data set will be described in detail in the forthcoming \cite{For+13}.

$^{2}$ New analysis of this data is presented here.

$^{3}$ 3 galaxies from this sample were not included as they overlap with the SINS/zC-SINF sample.

$^{4}$ These galaxies were selected from the total sample such that they had kinematic information based on the \ha \ls line from either H- or K-band, as the AO system on the VLT does not work as well in the J-band.

\end{table*}

Figure 1 shows the location of our entire sample in the SFR-\mstar \ls plane. As can be seen from the left panel, SFGs are well-sampled from a SFR of 10 to 500 \msun/yr and over 2 decades of stellar mass, from $\sim$2x10$^{9}$ to 2x10$^{11}$ \msun, and from the right panel, we see that both rotation and dispersion-dominated galaxies are widely distributed.

\begin{figure*}[tb]
\centerline{
\includegraphics[width=7in]{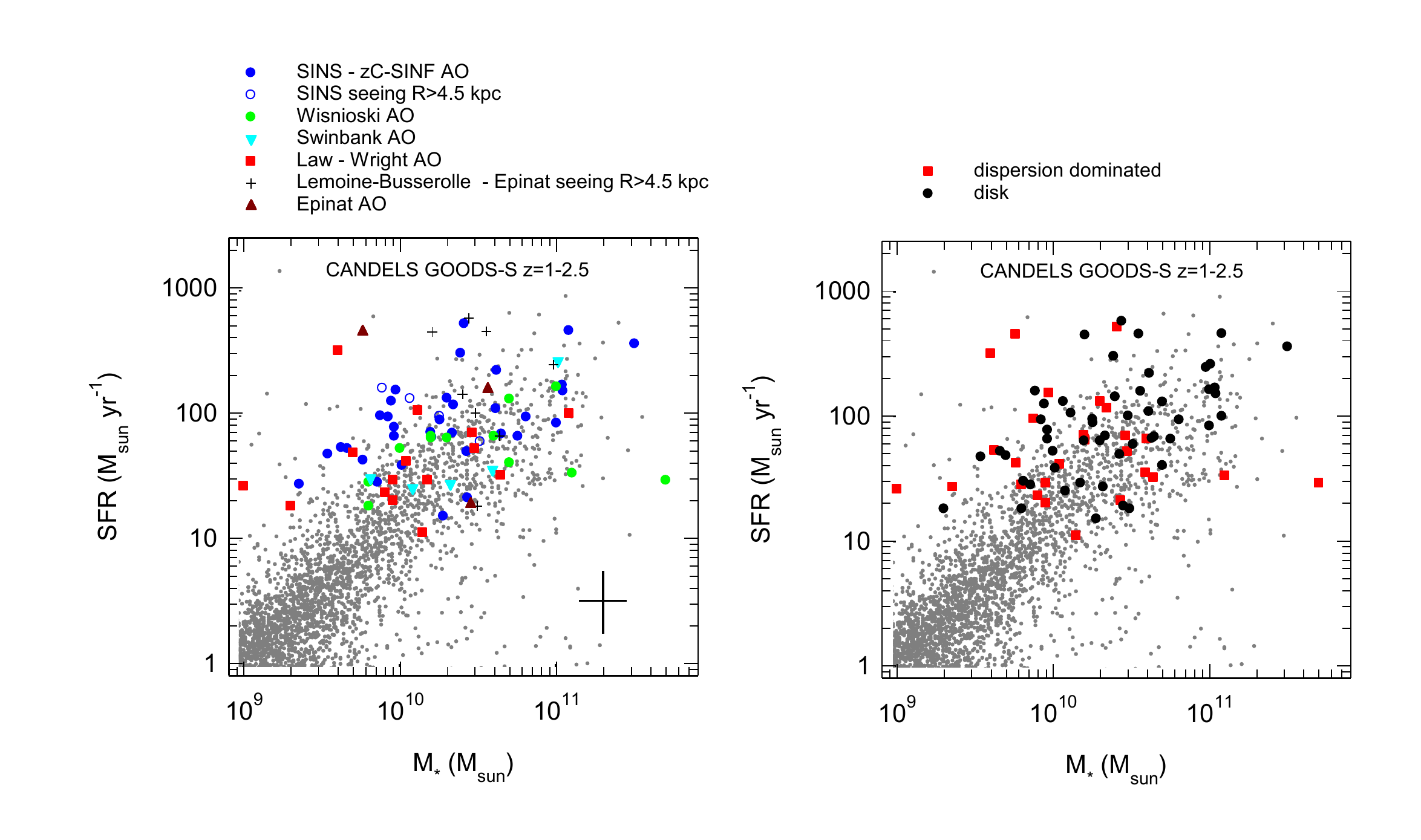}}
\caption{The star-formation rate/stellar mass plane. The grey points are from z = 1--2.5 galaxies from the CANDELS survey in the GOODS-S field. The left panel shows the distribution of galaxies from different parent samples used in our analysis. The blue closed (and open) circles are from the SINS/zC-SINF galaxies presented in this paper observed in AO (seeing-limited) mode. The red squares are from \cite{Law+09,Law+12c,Wri+09}, the cyan inverted triangles are from the SHiZELS survey presented in \cite{Swi+12a,Swi+12b}, the maroon triangles denote the AO galaxies from \cite{Epi+12}, the green circles represent the AO data from the WiggleZ survey \citep{Wis+11}, and the black `+'s denote galaxies from \cite{Epi+12} and \cite{LemLam10} that have \rhalf \textgreater 4.5 kpc and were observed in seeing-limited mode. The right panel shows the same distribution of galaxies according to their identification as dispersion- or rotation-dominated. Here, black circles represent galaxies determined to be rotation dominated, and red squares denote galaxies determined to be dispersion-dominated. Note that wherever possible, we used averages of the H$\alpha$- and SED-derived SFRs.}
\end{figure*}
  
\subsection{Determination of Galaxy Properties for the SINS/zC-SINF sample}

Stellar masses, star formation rates and stellar ages are derived from SED modeling of broad-band photometry in \cite{For+09}, \cite{Man+11} and \cite{For+11a} and assume either constant star formation histories or exponentially declining models with \cite{BruCha03} tracks. The ages are highly uncertain and should be interpreted in the sense that the bulk of galaxy light comes from stars younger than this ``age'', and as such are best used as relative and not absolute ages. The star formation rates are taken as an average of those calculated from SED modeling and \ha-based star formation rates, derived using SFR=L(\ha)/(2.1x10$^{41}$ erg/s) (\msun yr$^{-1}$) \citep[corrected for a \cite{Cha03} IMF]{Ken98}, where L(\ha) is extinction corrected based on the \cite{Cal+00} reddening law with A$_{V,gas}$=2.3xA$_{V,SED}$. \cite{For+09} and \cite{Wuy+11a} have shown that an extra attenuation factor of $\sim$2.3 is a good representation of the global extinction towards HII regions relative to the bulk of starlight for \ztwo SFGs.

We calculate molecular gas masses (M$_{gas}$) and surface densities ($\Sigma_{gas}$= 0.5xM$_{gas}$/($\pi$(\rhalf)$^{2}$) from the \ha- and SED-derived SFRs using the star formation rate/molecular gas mass relation calibrated from the IRAM Large Program of CO in z$\sim$1--2 SFGs \citep[$t_{depl} = 1.5 \times (1+z)^{-1}$ Gyr,][]{Tac+12}. This simple linear relation has been shown to hold for both local and high-z SFGs, yielding \mgas(\msun) = $t_{depl} ~\times$ SFR (\msun/yr) \citep{Big+08,Tac+10,Tac+12,Sai+12} down to $\sim$1 kpc scales, with scatter in the relation of $\sim$0.3 dex due to the variation in slope and normalization from various studies and also intrinsic scatter in the relation \citep[see:][]{Gen+10,Dad+10,Tac+12}. The dynamical mass is calculated as, M$_{dyn}$ = 2\rhalf(v$_{rot}^{2}+3.4 \times \sigma_{0}^{2}$)/G, where \vrot \ls is the inclination-corrected rotation velocity and \sn \ls is the intrinsic galaxy dispersion. The factor of 3.4 accounts for pressure support (asymmetric drift) in an exponential distribution.

Inclinations were determined from the minor axis to major axis ratio of the \ha surface brightness distribution. We found overall good agreement between the morphological axes and those determined from kinematics, however smaller objects tended to exhibit a large scatter in the position angles determined from the two methods. In addition, some of the larger, more face-on systems have their axis ratios and position angles partially affected by small-scale regions of enhanced surface brightness. Following \cite{Law+12a}, we considered the intrinsic z-thickness of high-z SFGs in deriving sin(i) (assuming \textless z\textgreater/\textless R\textgreater=0.2), although the inferred inclinations are not significantly different from those in the thin disk approximation. This technique assumes that the light distribution is intrinsically circularly symmetric. However, \cite{Law+12a} find fewer b/a$\sim$1 systems than expected for a population of randomly oriented inclined disks. Wuyts et al. (in prep.) find the same trend at all z from 0 to 2, although with lower amplitude at lower redshift. Independent of whether this deviation from circular symmetry in the light distribution is due to an intrinsically triaxial light or matter distribution or whether it's due to the distribution of clumps distorting the symmetry, this results in additional uncertainties in our derived inclinations, mainly for face-on systems.

Intrinsic (corrected for the spatial resolution) half-light radii are based on 2D exponential profile (i.e. Sersic profile with index n = 1) fits to the \ha surface brightness distributions using the code GALFIT \citep{Pen+02} following the methodology described by \cite{For+11a} (details for the present SINS/zC-SINF AO sample are given by \cite{For+13}).  For six cases (BX389, BX482, GMASS-2540, ZC405501, ZC406690, and ZC410041), a Gaussian profile (Sersic n = 0.5) was found to provide a significantly better representation of the data and we adopted the sizes derived from these Gaussian fits. Sersic profiles with n $\lesssim$ 1 are motivated by the previous analysis of a subset of six SINS galaxies \citep{For+11a}, and the more recent result from a larger sample of z $\sim$ 1.5 SFGs by \cite{Nel+12b}.

The oxygen abundances are derived from the observed \niiha \ls flux ratio using the \cite{PetPag04} calibration: 12+log$_{10}$(O/H) = 8.90 + 0.57 $\times$ log$_{10}$(\niiha). This calibration is most often used for high-z galaxies with emission line data of only \ha and \nii, and thus we choose it for simplicity of comparison with previous work. We note that there are several uncertainties associated with this particular calibration \citep[see e.g.,][]{KewEll08}, so it is taken as a relative measure of gas-phase abundance among our objects.

\subsection{Determination of Kinematic Properties}

For each galaxy, we computed the observed velocity gradient ($\Delta v_{grad}$) as the maximum velocity difference across the source from the velocity map produced from the data cube. The total Gaussian line width integrated over the source ($\sigma_{tot}$) is calculated from the width of the \ha line from the spatially integrated spectrum for each galaxy. For SFGs with a substantial component of rotation or large-scale orbital motion, $\sigma_{tot}$ is strongly affected or even dominated by beam-smeared rotation. However, all of the SINS/zC-SINF galaxies presented in this paper are resolved well enough in either the AO or seeing-limited data to estimate the average intrinsic local velocity dispersion (\sn) from the velocity dispersion maps, and we minimize the effects of beam smearing by measuring the observed velocity dispersion in the outer parts of the source. Note that all quoted velocity dispersions have been corrected for the instrumental response function and are thus intrinsic quantities of the source. We determine the rotational velocity (v$_{rot}$) by correcting $\Delta v_{grad}$ for our best-fit inclination. 

In determining the kinematic properties for the other galaxies in our larger sample, we attempt to use as consistent a method as possible. \cite{Epi+12} and \cite{LemLam10} use a similar method to ours for determining \vrot \ls and \sn, so we use the values presented in their papers. As with our method, they create maps of the velocity and dispersion fields for their galaxies that are corrected for instrumental resolution and additionally subtract (in quadrature) a beam-smearing component as estimated from thin rotating-disk models. \cite{Dav+11} has shown that deriving kinematic parameters from such modeling has the advantage of not being systematically biased by beam smearing and most often produces more correct results. However, there can be some inaccuracy in the resulting parameters due to low S/N, and \cite{Dav+11} suggest that extracting \sn \ls from the outer regions of the galaxy (as we do with the SINS/zC-SINF galaxies) could help alleviate this issue. Either way, these two techniques should provide similar results. For the SHiZELS galaxies from \cite{Swi+12a}, we take \vrot \ls and \sn \ls from their Figure 4, assuming the values in the outer regions of the galaxies, for consistency with our method. We note the the values they give for $\sigma$ are almost all larger than the ones we assume. This is consistent with the bias towards higher \sn \ls values demonstrated by \cite{Dav+11}, which is found when correcting for beam smearing without generating disk models or without ignoring the central regions of the galaxy. We do the same for the \cite{Wri+09} galaxies, using their velocity and velocity dispersion maps. For the \cite{Law+09,Law+12c} galaxies, we take their \vrot \ls and \sn \ls values, which were derived from fitting a Gaussian to the line profile from each pixel and flux-weighting to determine an average. As mentioned in their work and \cite{Dav+11}, this produces an upper-limit for \sn \ls once beam-smearing is accounted for, and this is reflected in Figures 5 through 9. \cite{Wis+11} do not derive \sn \ls values in a way that is comparable to our values, so we use their dataset for \vrot \ls comparison only.

We follow \cite{For+09} and operationally define dispersion dominated and rotation dominated SFGs as having $\Delta v_{grad}/2\sigma_{tot}$ \textless0.4 and $\geq$0.4, respectively, for both AO and seeing-limited data. This operational divide is based on disk models of masses and sizes roughly spanning the range observed in our sample and with typical beam-smearing of seeing-limited data. Once corrected for inclination and beam smearing, it closely corresponds to the more intuitive physical definition \vrot/\sn \textless1 and $\geq$1. None of the galaxies in our sample are obvious major mergers but several may be minor mergers with a kinematically related companion galaxy within 10--20 kpc. For these, we consider the kinematic properties of the entire system.

We caution that a detection of a velocity gradient is certainly required but is by itself not sufficient to prove that a galaxy is rotating. If the velocity gradient is along the morphological major axis and at the same time, a peak of velocity dispersion is present near the morphological centroid, the evidence becomes more convincing. This second line of evidence is present for most of the galaxies in the SINS/zC-SINF sample. The ultimate, but hardest, criterion is the detection of a symmetric `spider diagram' pattern in the iso-velocity contours whose shape/curvature is consistent with the inclination estimated from the minor to major axis ratio \citep{vdKruAll78}. This most demanding proof is not reached by any of the compact dispersion dominated systems, and only by very few of the large disks/rings.
 
\begin{figure*}[tb]
\centerline{
\includegraphics[width=7in]{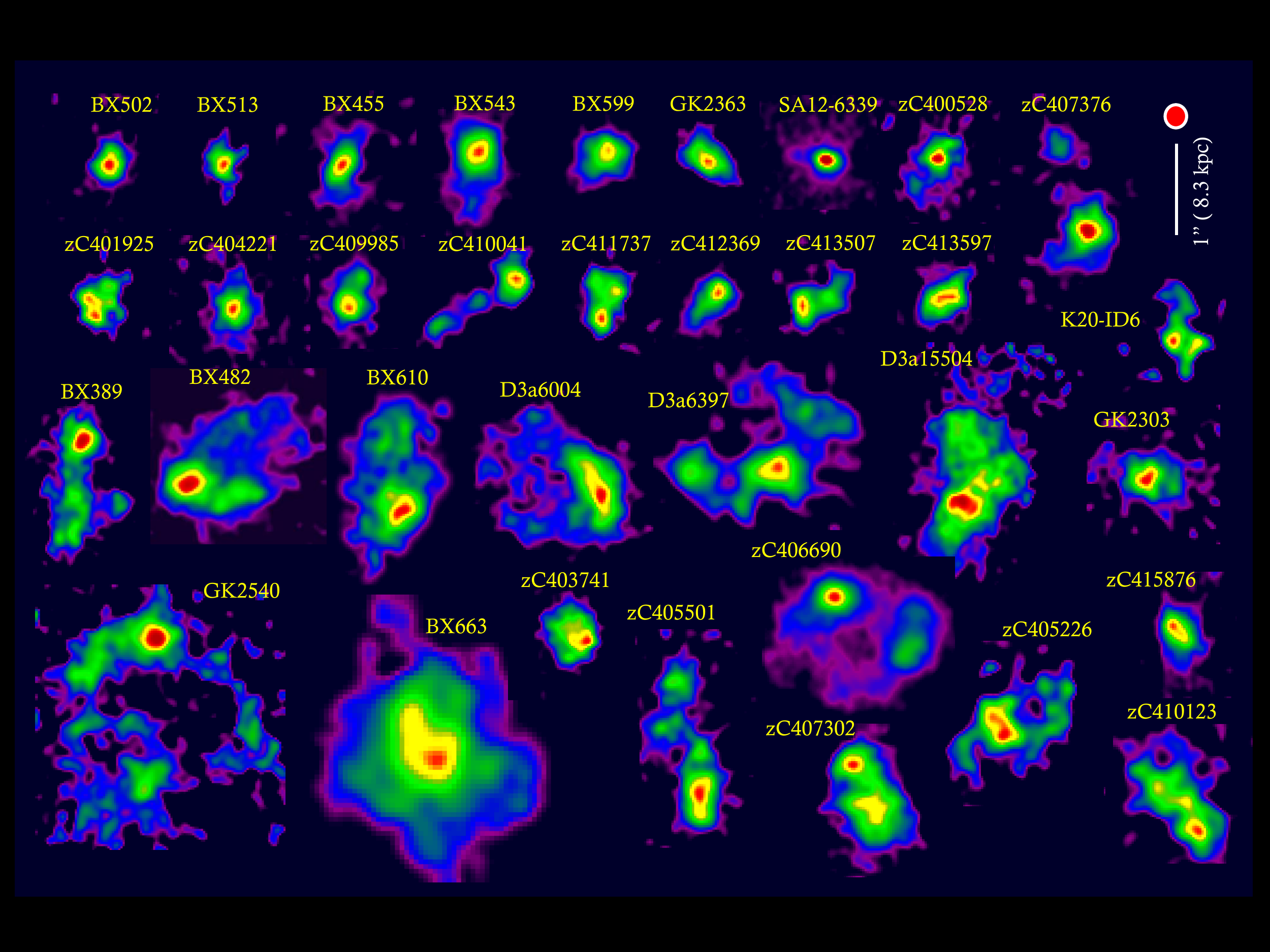}}
\caption{Integrated \ha maps in AO mode (0.2--0.3'' FWHM) of the 34 SFGs in our (AO) sample. The top two rows contain the dispersion dominated SFGs (as defined by the criterion ($\Delta v_{grad}$/(2x$\sigma_{tot}))_{seeing} \leq 0.4$, \cite{For+09}, and the rest are rotation dominated. The maps were interpolated on a pixel scale of 0.025'' and are all plotted on the same angular scale. The typical FWHM resolution is shown as a red circle. The dispersion dominated galaxies tend to be more compact than the rotation dominated galaxies with the peak of their \ha emission in the center.}
\end{figure*}

\section{The Nature of Dispersion Dominated SFGs}

As discussed in the Introduction, all investigations of the spatially resolved ionized gas dynamics have consistently found that z$\sim$1--3 SFGs possess large internal random motions, in addition to large scale ordered velocities, such as rotation (in a disk), or orbital motion (in an interacting or merging system).  In a significant fraction (anywhere from 20-90\%, depending on the sample and definition of kinematic parameters) of the available IFU samples, the random motion component even appears to dominate, similar to what is seen in spheroidal galaxies. This characteristic feature of high-z galaxies is surprising, since in a gas rich system, supersonic turbulence should dissipate on a dynamical time scale ($\sim10^{7-7.5}$ yr), unless it is continuously replenished. 

\subsection{Dispersion Dominated Galaxies are Small}

Figure 2 shows postage stamps of the velocity-integrated \ha emission maps of 34 of the SINS and zC-SINF AO data sets. The galaxies are ordered such that the 17 galaxies in the top two rows have $\Delta v_{grad}$/2$\sigma_{tot}$ below or very near 0.4, and are thus classified as `dispersion dominated' based on the seeing limited data of  \cite{For+09,Man+11}.  According to the same criterion, the other 17 galaxies in Figure 2 are rotationally supported. None of the 34 is an obvious major merger, but the presence of a spatially well-separated secondary component in six of these SFGs (BX543, BX513, zC409985, zC410041, zC407376, zC407302) may be evidence for an ongoing minor merger (mass ratio \textgreater3:1), although in some cases it is difficult with the data in hand to distinguish unambiguously between bright clumps in large disks and minor merger systems.

Figure 2 immediately shows that the most obvious characteristic of dispersion dominated systems is their small size. If we look at those galaxies with \rhalf$\leq$3 kpc, then most of these 13 smaller systems fulfill one or several of the kinematic definitions of dispersion dominated galaxies as introduced in Section 2.4,

\begin{equation}
(\Delta v_{grad}/(2 \times \sigma_{tot}))_{seeing} \leq 0.4
\end{equation}
\begin{equation}
(\Delta v_{grad}/(2 \times \sigma_{tot}))_{AO} \leq 0.4 \\
\end{equation}
\begin{equation}
v_{rot}/\sigma_{0} \leq 1
\end{equation}

Comparison of seeing limited (FWHM $\sim$0.56'') and AO-scale (FWHM $\sim$0.20'') data of the same galaxies shows that the classification as dispersion or rotation dominated can depend strongly on angular resolution. This is demonstrated in Figures 3 and 4, which show the velocity and velocity dispersion fields for Q1623-BX455 and GMASS-2363 in both seeing and AO modes. As the galaxies are small (\rhalf$\sim$0.2''), the seeing limited data (resolution $\sim$ 0.5'') do not resolve them and appear to confirm their dispersion dominated classification. However, with the AO data (resolution $\sim$ 0.2''), both galaxies appear to be inclined rotating disks, based on the first two criteria for rotation outlined at the end of section 2 (velocity gradient along the morphological major axis and peak in velocity dispersion near the morphological center), yet like most high-z galaxies, neither exhibit the `spider diagram' required for a more definitive proof.

\begin{figure}[tb]
\centerline{
\includegraphics[width=3.5in]{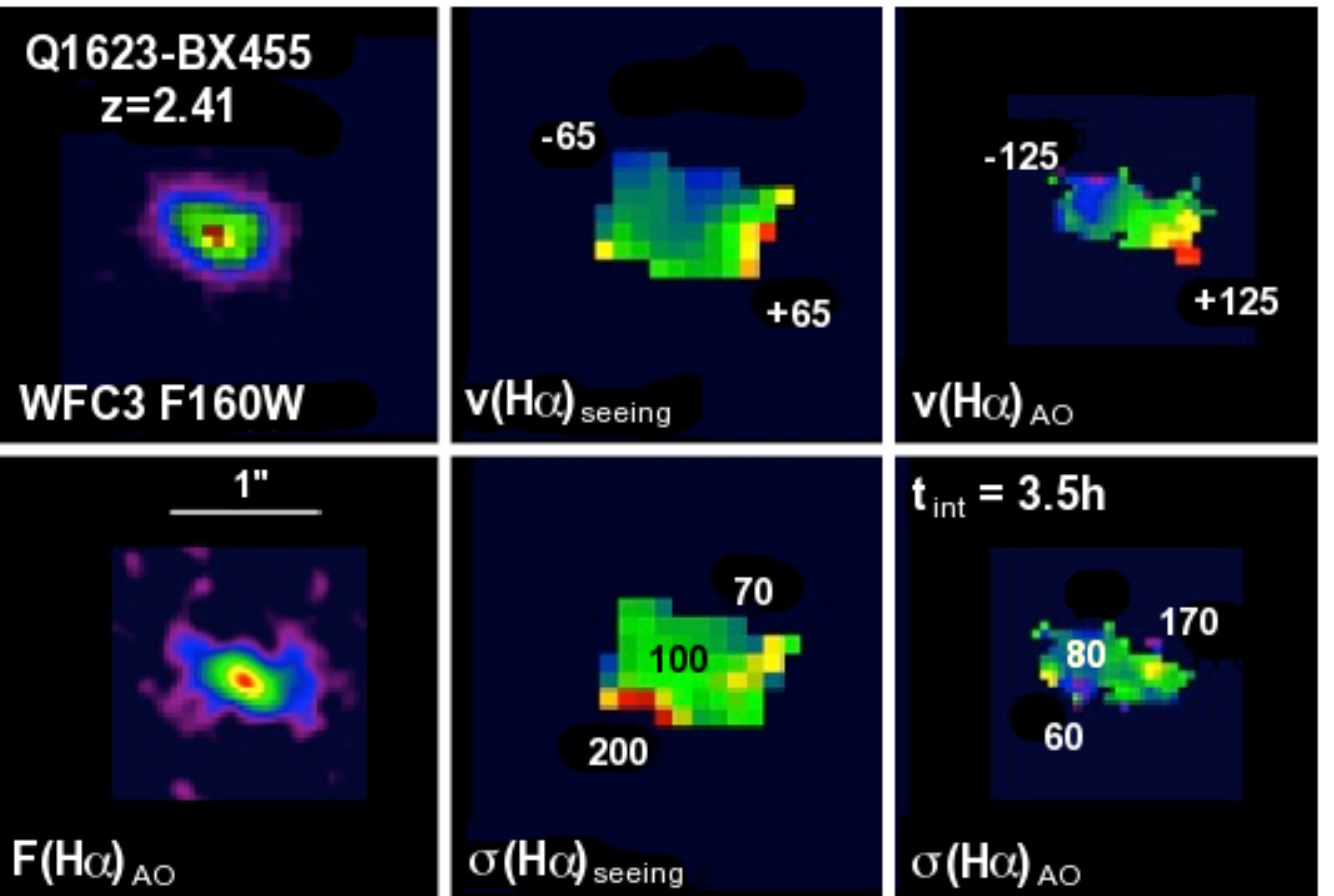}}
\caption{Top row, left to right: HST WFC3 \citep{Law+12a} image, seeing-limited velocity field, AO velocity field. Bottom row, left to right: \ha AO image, seeing-limited velocity dispersion field and AO velocity dispersion field of Q1623-BX455 \citep{For+09}. Much of the rotation apparent from the AO velocity field is beam-smeared out with the seeing-limited data.}
\end{figure}

\begin{figure}[tb]
\centerline{
\includegraphics[width=3.5in]{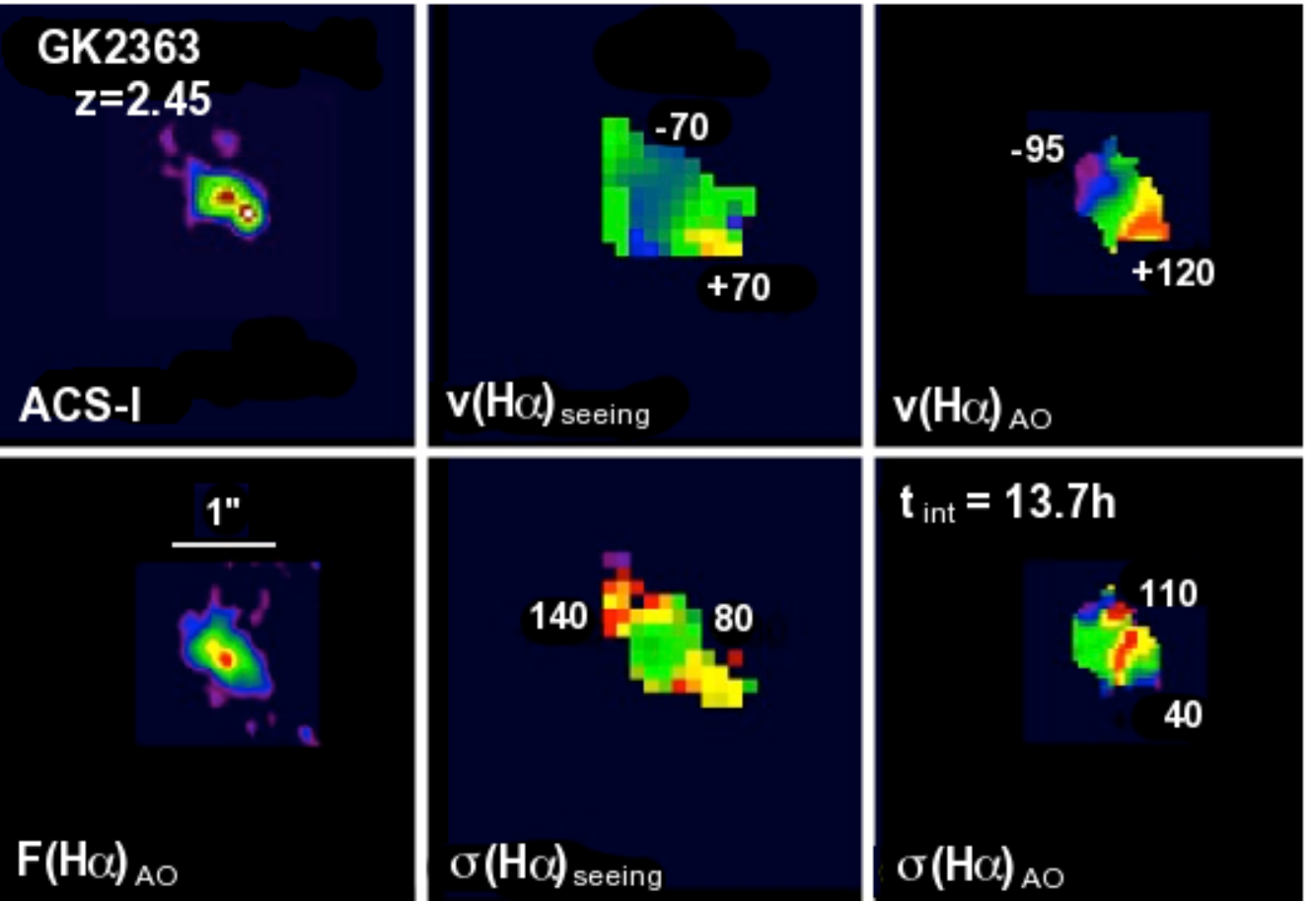}}
\caption{Top row, left to right: HST ACS I-band image, seeing-limited velocity field, AO velocity field. Bottom row, left to right: \ha AO image, seeing-limited velocity dispersion field and AO velocity dispersion field of GMASS-2363 \citep{For+09}.}
\end{figure}

\subsection{Impact of Resolution on Kinematic Classification}

\begin{figure*}[tb]
\centerline{
\includegraphics[width=6in]{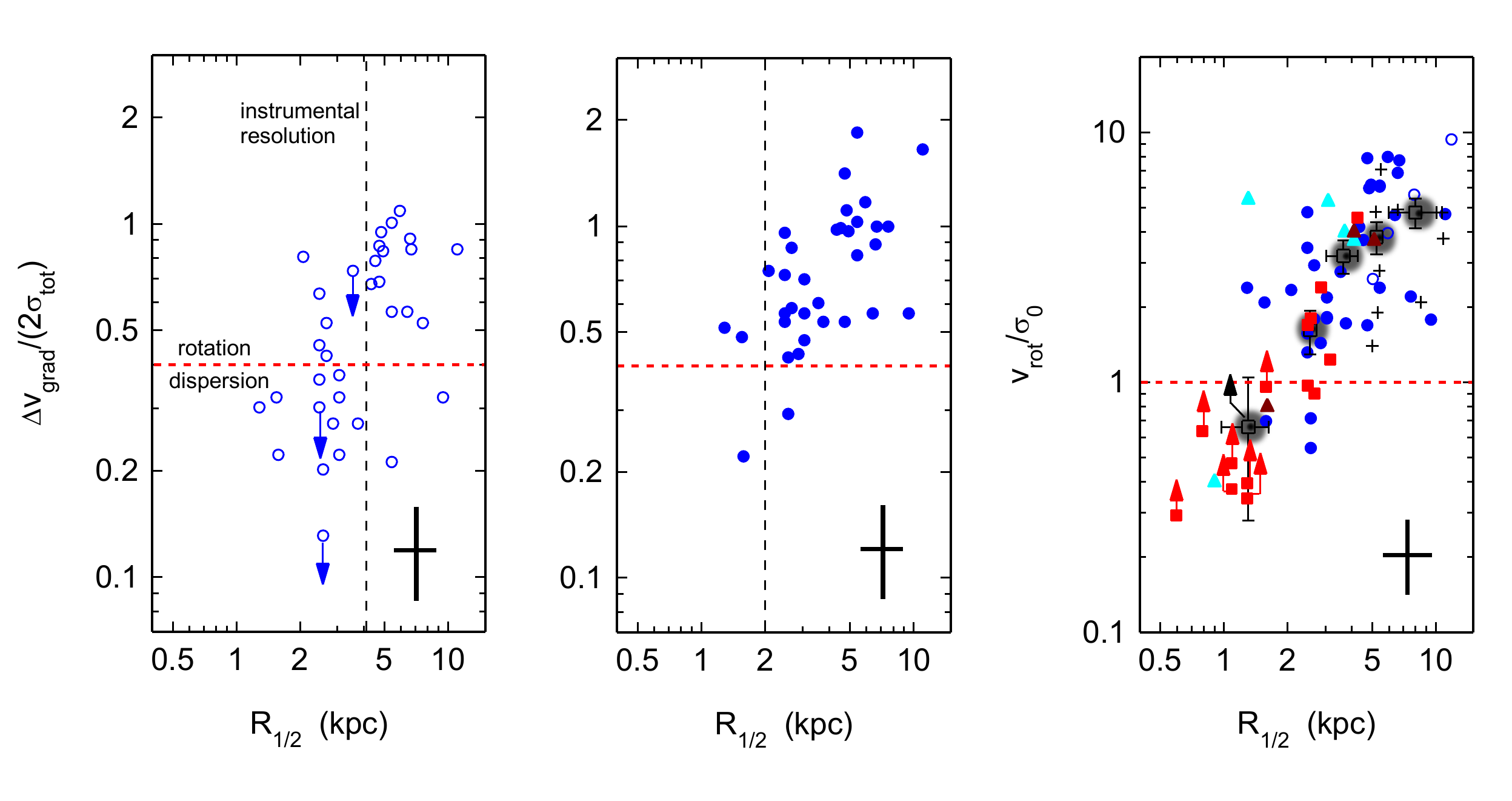}}
\caption{ Dependence of the ratio of rotation to dispersion on the half light source radius (\rhalf) and resolution. Left panel: seeing-limited SINS/zC-SINF data \citep{For+09,Man+11}. Middle panel: AO SINS/zC-SINF data \citep[this paper and][]{For+13}. Right panel: Combined AO and seeing-limited data for galaxies that have sufficient angular resolution for dynamical modeling to obtain the inclination corrected v$_{rot}$ and the beam smearing corrected $\sigma_{0}$ (i.e. all galaxies with AO data and seeing-limited only galaxies with \rhalf \textgreater 4.5 kpc). The colored symbols are the same as for the left panel of Figure 1. Grey filled circles denote the median-binned values in four radius bins with horizontal error bars representing the standard deviation in the bins and vertical error bars representing the 1$\sigma$ uncertainty. Many of the same galaxies are plotted in all three panels (for the SINS data). The horizontal dashed red lines mark the operational divide between dispersion and rotation dominated SFGs for the corresponding criteria for each panel. The vertical dashed black lines in the left and middle panels mark the FWHM spatial resolution of the data. Typical 1$\sigma$ error bars for individual measurements are plotted at the bottom of each panel. Many of the galaxies characterized as dispersion dominated according to the seeing-limited data (left panel), would be considered rotation dominated with the higher-resolution AO data (middle panel) or with full dynamical modeling (right panel), and an important factor in the kinematic classification is the ratio of the galaxy size to the instrumental resolution. }
\end{figure*}

In order to explore the impact of instrumental resolution on the classification (scenario 2 from Section I) of galaxies as dispersion- or rotation-dominated, we used the sample of 34 SINS/zC-SINF galaxies, for which we have both seeing limited and AO resolution data, taken with the same instrument and analyzed with the same tools. We compare the location of these galaxies in the $\Delta v_{grad}/(2 \times \sigma_{tot})$ vs. \rhalf \ls plane in Figure 5 for both seeing-limited data (left panel) and AO data (middle panel). The shift to higher $\Delta v_{grad}/(2 \times \sigma_{tot})$, and thus more rotation-dominated classification, with higher resolution data is clear. We also show our entire sample in the \vrot/\sn \ls versus \rhalf \ls plane in the right panel. 

It is apparent that for all IFU data sets the dispersion dominated classification correlates with the intrinsic source size (smaller galaxies are more likely dispersion dominated), although they are not perfectly matched. But as we see from the first two panels, this classification also depends on the ratio of resolution to source size, such that poorly resolved galaxies are very likely classified as dispersion dominated. Given the criteria above, formally 41\% of the 34 SINS/zC-SINF galaxies are classified as dispersion dominated based on seeing limited data. That fraction drops to 6--9\% for the same galaxies using AO data. If we include 1$\sigma$ error bars, up to 59\% of the galaxies could be classified as dispersion dominated with seeing limited data and less than 35\% with AO data.

In turn, the strong majority of the SINS/zC-SINF SFGs observed with AO can then be characterized by a velocity gradient along a single axis that is identical with or close to the morphological major axis. This suggests that rotation dominates the larger scale velocity field, although the observed pattern could also be matched by orbital motion in a binary minor merger in a few cases. However, we do not detect symmetric double light distributions or velocity reversals in any of the SFGs in our sample, which would be indicative of a major merger.

The empirical assessment drawn from Figure 5 is supported by creating simple toy models of turbulent but rotationally supported disks with intrinsic \vrot/\sn$\sim$1--5. We ``observe'' model disks with varying sizes, masses and inclinations with seeing and AO scale resolutions and analyze them in the same way as our SINS/zC-SINF data. Their location in the empirical $\Delta v_{grad}/2\sigma_{tot}$ -- \rhalf \ls and \vrot/\sn -- \rhalf \ls planes overlaps with the majority of the data. A fraction of the model disks indeed come to reside in the locus of `dispersion dominated' galaxies, although they are intrinsically rotationally supported.

Finally, we carry out weighted linear fits to the data in log-log space (i.e. log(\vrot/\sn) \ls vs. log(\rhalf)) and calculate the formal significance of the slope differing from 0. The resulting best-fit slope and significance for data in the right panel of Figure 5 (and the remaining Figures) is presented in Table 2.

\begin{table}
\caption{Significance of Correlation with Kinematic Parameters}
\scriptsize{
\renewcommand{\tabcolsep}{2pt}
\begin{tabular}{l l c c c}
\\
\hline
\hline 
\textbf{Parameter 1} & \textbf{Param. 2} &  \textbf{Lin. Slope of Fit} & \textbf{Sign.$^{1}$} & \textbf{See Fig.:} \vspace{2pt} \\
\hline 
log(\vrot/\sn) & log(\rhalf) & 0.77 $\pm$ 0.12 & 6.4 & 5 \\
log(\vrot) & log(\rhalf) & 0.62 $\pm$ 0.094 & 6.6 & 6 \\
log(\sn) & log(\rhalf) & -0.14 $\pm$ 0.05 & 2.8 & 6 \\
log(M$_{dyn}$) & log(\vrot/\sn) & 1.19 $\pm$ 0.097 & 12.3 & 7\\
log(M$_{*}$) & log(\vrot/\sn) & 0.54 $\pm$ 0.14 & 3.9 & 7 \\
log(f$_{gas}$) & log(\vrot/\sn) & -0.028 $\pm$ 0.05 & 0.56 & 8\\
12+log(O/H) & log(\vrot/\sn) & 0.1 $\pm$ 0.046 & 2.2 & 8 \\
log($\Sigma_{SFR}$) & log(\vrot/\sn) & -0.89 $\pm$ 0.18 & 4.9 & 9 \\
\hline

\end{tabular}}

$^{1}$ Significance of best fit linear slope as opposed to slope of 0.
\end{table}

A similar dependence of kinematic classification on resolution was discussed in \cite{Jon+10} \citep[see also:][]{Gon+10}. They found that without the boost in spatial resolution provided by gravitational lensing, many of the galaxies that they identified as rotating disks would not be distinguishable from mergers or dispersion dominated sources based on their velocity profiles, even with the use of AO. However, we caution that our sources are generally much larger than those in the \cite{Jon+10} work (\textless \rhalf \textgreater \ls $\sim$ 2 kpc), and will thus suffer less beam smearing with the use of AO even without the added spatial resolution of gravitational lensing.

\subsection{Why Are Small Galaxies More Likely to be Dispersion Dominated?}

\begin{figure*}[tb]
\centerline{
\includegraphics[width=6in]{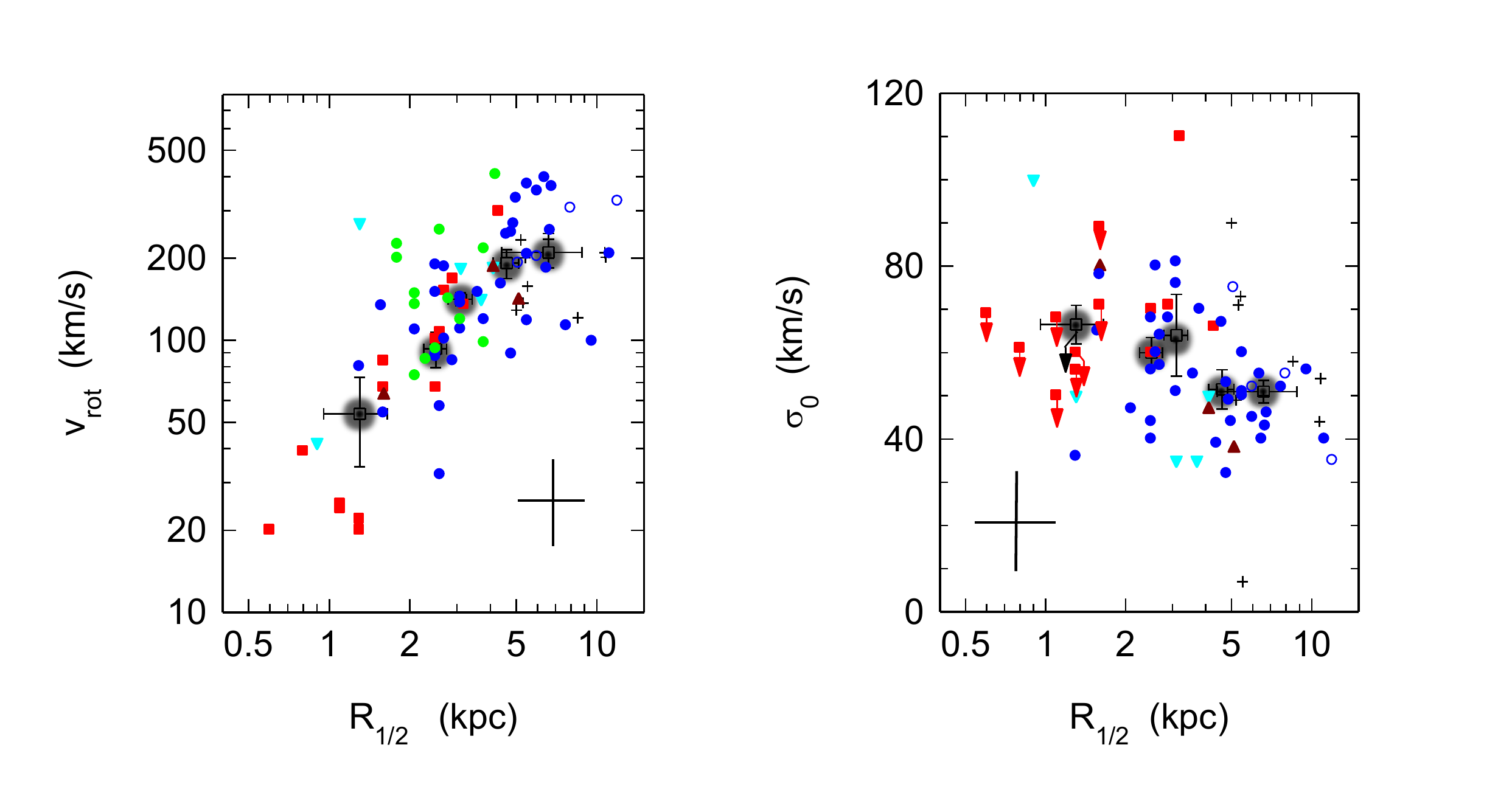}}
\caption{Dependence of \vrot \ls (left panel) and \sn \ls (right panel) on \rhalf. The data symbols are the same as in Figures 1 and 5. The strong trend in the left panel can be well fit by a linear relation: log(\vrot)= 0.62 ($\pm$0.094) $\times$ log(\rhalf) + 1.9 ($\pm$0.06) (see Table 2). In contrast, the right panel does not appear to show a significant trend, considering that many of the red data points (those from \cite{Law+09,Law+12c}) are strictly upper limits to \sn. Thus the trend for smaller galaxies to be dispersion dominated is in part due to the combination of a possible floor of velocity dispersion, and a linear increase of rotation velocity with size. We note that the characteristic error bars do not include potential additional uncertainties due to deviations from circular symmetry. However, these uncertainties really only strongly influence near face-on systems, which are very few in this data set.}
\end{figure*}

The last section has shown that resolution effects can make an intrinsically rotation dominated system appear to be dispersion dominated if it is small, especially with seeing limited data. However, the middle and right panels of Figure 5 show that $\Delta v_{grad}$/(2$\sigma_{tot}$) and \vrot/\sn \ls increase with radius even for well-resolved data sets, and even at AO resolution, there remain a number of dispersion dominated SFGs, for which the classification cannot be an instrumental effect. What causes this intrinsic dependence on size?

In Figure 6, we plot the rotation velocity (corrected for inclination) and the intrinsic velocity dispersion (corrected for beam-smearing) as a function of \rhalf. Rotation velocity increases strongly with \rhalf \ls but there is no strong trend with velocity dispersion and the running median suggests that \sn \ls is constant or perhaps slightly decreasing with \rhalf. The best fit weighted power law to the data in Figure 6 yields \vrot \ls = 59($\pm$13)$\times R^{0.73 (\pm0.18)}$, but a linear slope is also consistent with the data within the 2$\sigma$ fit error, which would be physically motivated by centrifugally supported baryonic disks of constant angular momentum parameter embedded in a virialized dark matter halo \citep{MoMaoWhi98}. Figure 6 thus shows that small galaxies are intrinsically more likely to be dispersion dominated because of the near-constant value of the local velocity dispersion in all high-z SFGs, which may suggest a velocity dispersion floor. Such a floor (\textless \sn \textgreater \ls  $\sim$ 60 $\pm$ 10 \kms) may be caused by star formation feedback or by the dissipation of gravitational energy within and at the outer edge of the disk \citep{Imm+04a,Imm+04b,Wes+07,BouElmMar09,Age+09,Cev+10,Gen+08,Gen+11,GenDekCac12, Gen+12, ElmBur10, KruBur10}.  

\subsection{Dependence on Stellar Mass, Gas Fraction and Age}

\begin{figure*}[tb]
\centerline{
\includegraphics[width=6in]{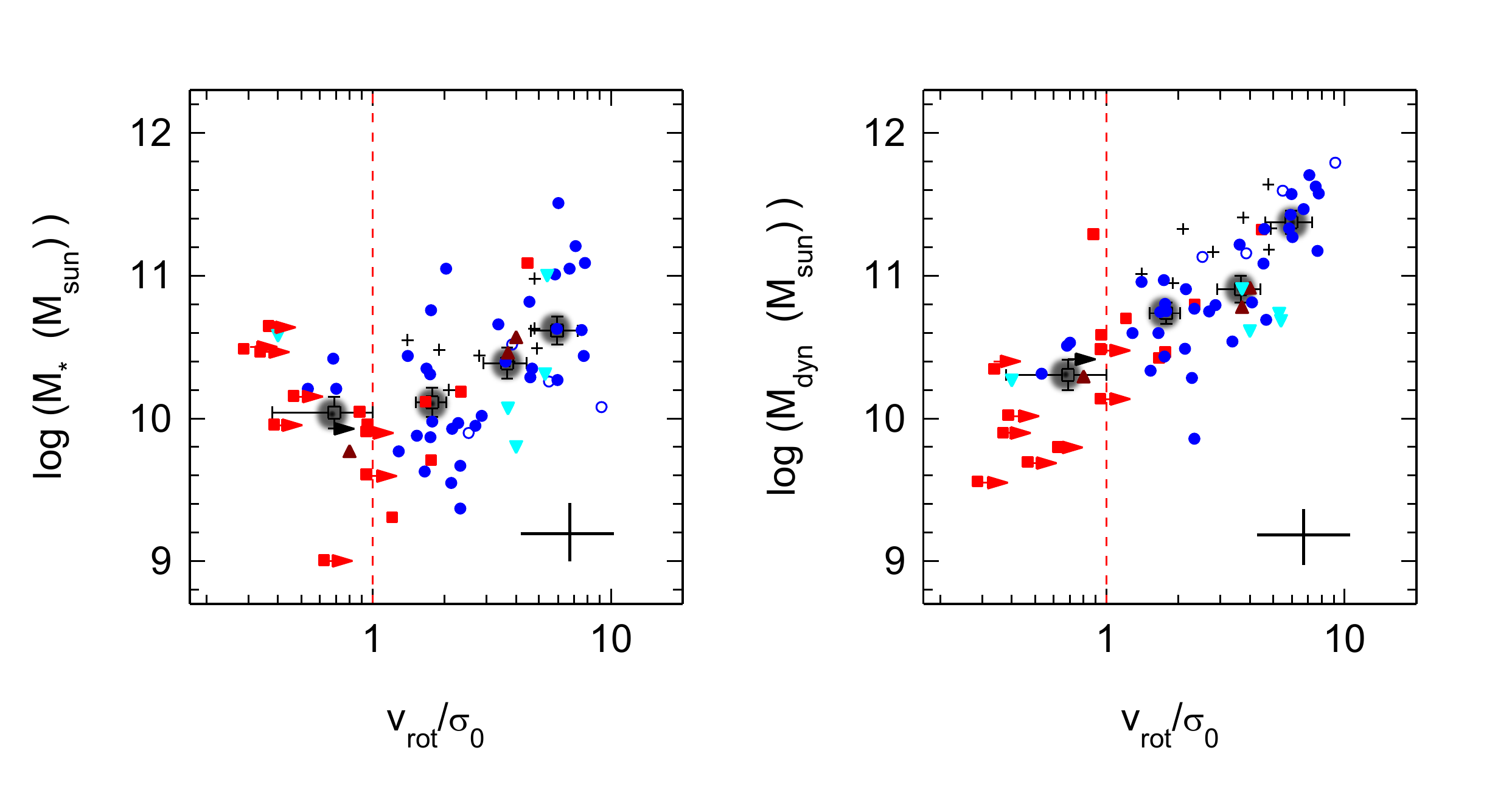}}
\caption{Dependence of stellar mass (left) and dynamical mass (right) on \vrot/\sn. Symbols are the same as in Figures 1, 5 and 6.}
\end{figure*}

\begin{figure*}[tb]
\centerline{
\includegraphics[width=5.5in]{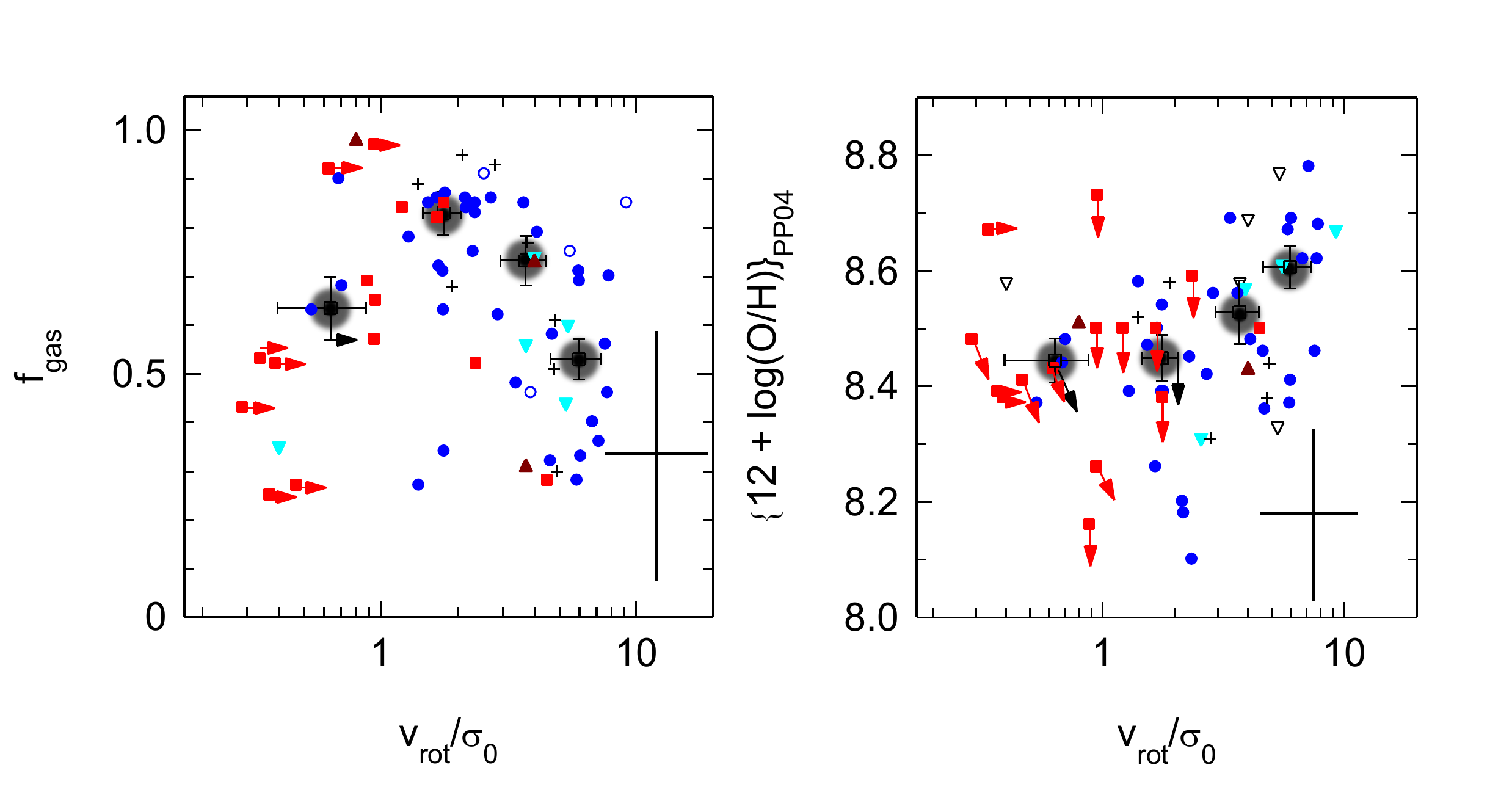}}
\caption{Dependence of baryonic gas fraction (f$_{gas}$=(M$_{mol-gas}$/(M$_{mol-gas}$+M$_{*}$)) (left panel) and gas phase oxygen abundance (right panel) on \vrot/\sn. Symbols are the same as in Figures 1 and 5 to 7. For the MASSIV sample, the gas masses and \ha half-light radii are listed in \cite{Ver+12} and the metallicities in \cite{Que+12}. Gas fraction and metallicity are correlated with kinematical type, such that dispersion dominated galaxies are more gas rich and slightly more metal poor.}
\end{figure*}

Figure 7 shows that dispersion dominated galaxies tend to have smaller stellar and dynamical masses, in agreement with the earlier conclusions of \cite{For+09,Law+09,Wis+11,Epi+12}. From the right panel of Figure 1, we see that while the dispersion dominated population (on the AO scale) is mostly found toward lower stellar masses, there is still substantial overlap with the location of disks. Thus, it seems that the best separation between the two kinematic populations is in terms of size and dynamical mass.

Figure 8 shows that the gas fraction (f$_{gas}$=(M$_{mol-gas}$/(M$_{mol-gas}$+M$_{*}$)) may be correlated with \vrot/\sn \ls, but that this trend becomes far more uncertain for the dispersion dominated objects. While there is a large range in gas fraction for the dispersion dominated objects, some have gas fractions as large as 90\%, much higher than for most of the larger rotating disks. These gas fractions are larger than those found by \cite{Tac+12} based on CO observations from Plateau de Bure for a sample of massive (\mstar \textgreater 3$\times 10^{10}$ \msun) SFGs at z = 1.2 and z = 2.2. This is likely because of a combination of the lower masses and above-main sequence location of many dispersion dominated galaxies (see right panel of Figure 1). \cite{Tac+12} have shown that gas fractions increase with decreasing stellar mass and with increasing offset from the star-forming `main sequence' in the stellar mass - SFR plane.

There is a modest difference in the strength of the \nii \ls line relative to the \ha \ls line between the dispersion and rotation-dominated sub-samples, implying a metallicity trend (see Figure 8). We stack the spectra of the SINS/zC-SINF galaxies by kinematic type, and find that the dispersion dominated sub-sample has \textless F\nii6584/F\ha\textgreater = 0.13 ($\pm$0.01), while the rotation dominated sub-sample has \textless F\nii6584/F\ha\textgreater = 0.19 ($\pm$0.01), consistent with the larger stellar masses of the rotation dominated sub-sample and the mass-metallicity relation \citep{Tre+04,Erb+06,Man+09}. The errors here derive from the uncertainty in the fitting of the emission lines. These averages do not include the galaxies identified to contain AGN.

\begin{figure}[tb]
\centerline{
\includegraphics[width=3.5in]{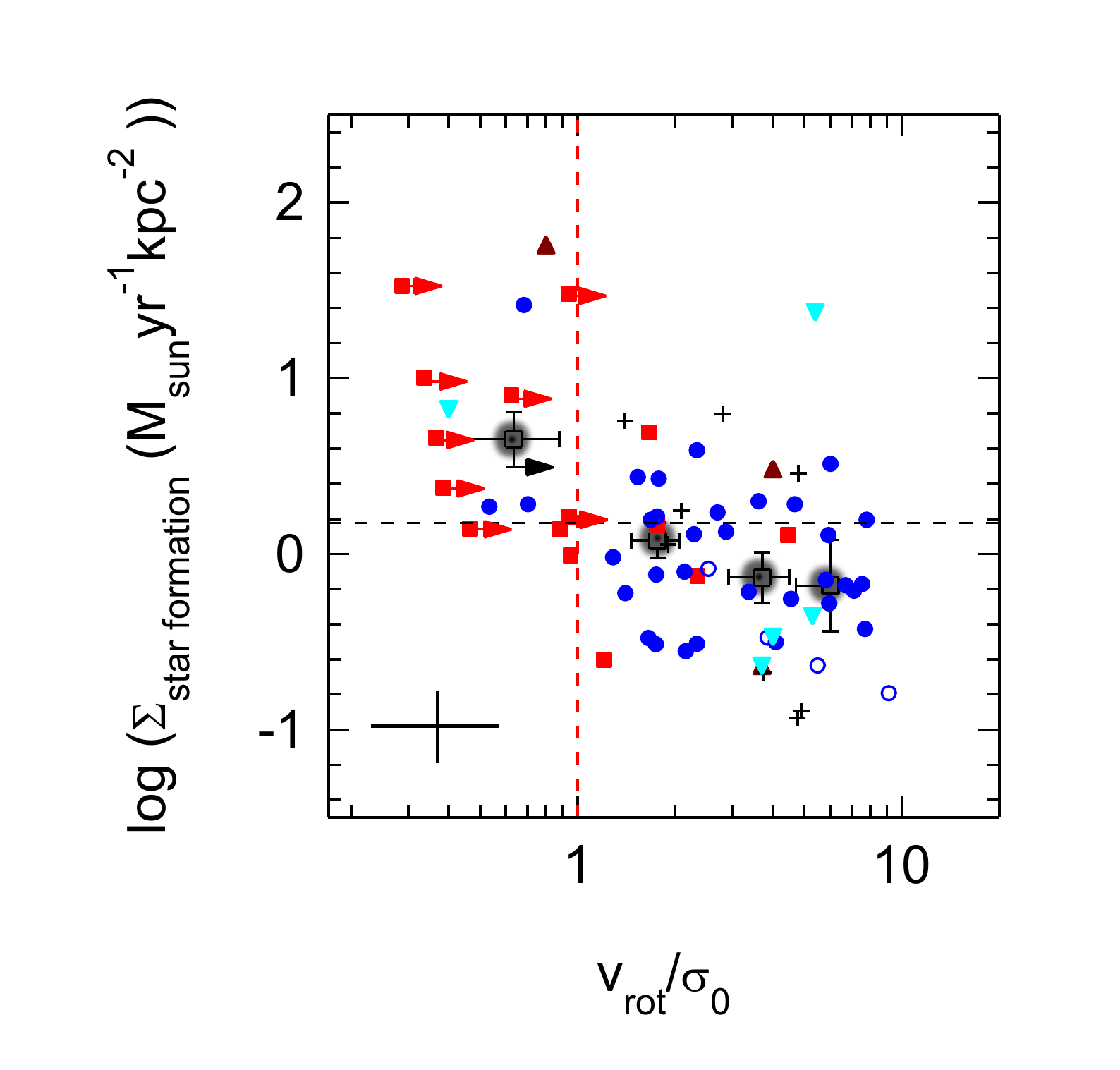}}
\caption{Dependence of star formation rate surface density (\sigsfr = 0.5 SFR/$\pi$(\rhalf)$^{2}$) on v$_{rot}/\sigma_{0}$. The symbols are the same as in Figures 1 and 5 to 8. The threshold for strong \ztwo galactic-scale outflows observed by \cite{New+12b} and marked by the black dashed line is at \sigsfr \ls $\sim$ 1.5 \sigsfrunits.}
\end{figure}

As found in \cite{New+12b}, z$\sim$2 SFGs with \sigsfr \textgreater 1 \sigsfrunits \ls have evidence for strong outflows, based on the ratio of broad \ha \ls emission to narrow \ha \ls emission (which likely traces star-forming gas in the disk). We find that dispersion-dominated galaxies generally have higher \sigsfr \ls than the sample as a whole (Figure 9). When coupled with Figures 6 and 7, this suggests that outflows may be stronger (or more likely) in galaxies with smaller rotation velocities and dynamical masses \citep[as suggested by theoretical models, e.g.][]{Mur+05}.

When comparing the mean ages of `rotation' or `dispersion dominated' galaxies from the SINS/zC-SINF sample we also see a relative trend. When classified using seeing-limited data, the dispersion dominated sample has a mean age of 460 Myr, while the rotation dominated galaxies have a mean age of 690 Myr. This contrast is even stronger when using AO data or the \vrot/\sn \ls criteria, wherein the mean ages are 120 and 650 Myr for the dispersion and rotation dominated samples, respectively. These ages pertain to the stars providing the bulk of the rest-frame UV and optical radiation, as these dominate the photometry used in our SED analysis.

We thus find that high-z SFGs may be classified as `dispersion dominated' for two main reasons, namely that (a) intrinsic rotation is beam-smeared by insufficient spatial resolution, and (b) the small galaxy size is accompanied by lower rotation velocity, which when coupled with the nearly constant floor of velocity dispersion leads to a low \vrot/\sn \ls ratio. These conclusions enable us to reflect on the question posed in the Introduction about what these objects really are. We find that:
\begin{itemize}
\item[] (1) The dispersion dominated objects may indeed be an earlier evolutionary stage of larger disks due to the lower stellar and dynamical masses, lower metallicities and possibly higher gas fractions. We discuss this possibility in more detail in the following section.
\item[] (2) It also appears that they are (smaller) rotating disks, with rotation that is masked by beam-smearing due to insufficient resolution in addition to intrinsically lower rotation velocities (likely due to their smaller dynamical masses).
\item[] (3) A fraction of the dispersion dominated systems may also be the product of a recent major merger. While the parent samples from which our SFGs were drawn already only contained a few binary systems with kinematic properties consistent with major mergers \citep[which we have intentionally excluded, see methods of e.g.][]{Sha+08,For+09}, and the new AO observations discussed here have not uncovered new major merger candidates of somewhat smaller size, there could still be late stage major mergers in our sample masquerading as dispersion dominated if they are sufficiently compact.
\item[] (4) It is less likely that the dispersion dominated objects are merely large clumps in low surface brightness disks. First, if this were the case, we would not expect to see more rotation with higher spatial resolution. In addition, when we stack all of the galaxies classified as dispersion dominated based on seeing-limited data, we find sersic indices of $\sim$ 0.5--1 with r$_{eff} \approx$ \textless \rhalf \textgreater, and without excess extended emission. \cite{Law+12a} also find that a stack of their `Type I' galaxies is well fit by an n $\sim$ 1 sersic profile, with r$_{eff}$ = 1.36 kpc. \\
\end{itemize}

\section{Implications for Galaxy Evolution}

Based on this combined sample of IFU datasets, we are able to hypothesize on how the dispersion dominated galaxies affect our picture of galaxy evolution. Our finding that most of these galaxies are both highly gas-rich and rotating (albeit smaller and more slowly than the larger disks) systems, naturally leads to an explanation of their formation in the context of gravitational instabilities in gas rich disks. If we frame the Toomre (or Jeans) mass in terms of gas fraction and v/$\sigma$, we find that for a marginally stable disk with Q$\sim$1, that R$_{toomre} \sim$ R$_{disk}$ $\times$ ($\sigma$/v) \citep{Gen+11}. \cite{Gen+11} used this argument to explain why \rhalf $\sim$ 5 kpc disks form $\sim$1 kpc clumps, however, this can also be used to show that for the dispersion-dominated galaxies (with \vrot/\sn $\sim$ 1), `clumps' will naturally form on the scale of \rhalf \ls (or~$\sim$2kpc). Thus in the same scenario in which clumps form in extended, gravitationally-unstable disks, the dispersion dominated galaxies will essentially form one giant highly-unstable clump.

In addition to the smaller sizes and potentially higher gas fractions, we also found that dispersion dominated galaxies tend to have lower stellar masses, younger ages (based on SED fitting), and lower inferred metallicities on average than rotation dominated galaxies. One can imagine a scenario in which the smaller dispersion-dominated galaxies are ``seeds'' for the larger and more massive rotating disks, and these larger galaxies at z $\sim$ 2.2 have merely evolved sooner. As these ``seeds'' continue to rapidly accrete gas, form stars and expel winds, they also grow in size, build up their stellar masses, increase in metallicity and decrease in gas fraction. We find that these dispersion dominated galaxies are more likely to have star formation surface densities above the wind `break-out' threshold proposed by \cite{New+12b} than their larger disky counterparts, implying that they drive outflows more efficiently than the latter ones do \citep[see also:][]{Law+12b}. The higher stellar masses will eventually stabilize the disks, and the larger sizes and increased rotation velocities will decrease the Toomre scale, shrinking the size of the star-forming regions (the clumps). A similar mechanism for the growth of dispersion-dominated galaxies into rotationally-supported disks was also suggested by \cite{Law+12b}. We note that our dispersion-dominated galaxies are on average not the same as the high-z compact star-forming galaxies (``blue nuggets'') recently reported on by \cite{Bar+13}, which they propose will soon quench and become compact quiescent galaxies. While two of the galaxies from our sample (BX502 and SA12-6339) do meet their mass/size criteria, the remainder of our smaller galaxies have much lower stellar masses (by a factor of $\sim$2--5).

The picture we have presented, in which stellar mass builds up in the centers of galaxies through this `compact dispersion-dominated' phase is supported by additional observational evidence. Based on 3D-HST \ha and rest-frame R-band data of z $\sim$ 1 SFGs, \cite{Nel+12} find that \ha emission is typically more extended than continuum emission, but that this is less often the case for the smallest objects (r$_{H\alpha}$ \textless 3 kpc) that have star formation surface densities \textgreater 1 \sigsfrunits, suggesting inside-out growth. Similarly, \cite{Wuy+12} find that z $\sim$ 1--2 SFGs from the CANDELS survey typically have large stellar bulges with high extinction and/or old stellar ages and UV-bright star-forming clumps with little or no excess in stellar mass at outer radii.

These unstable z $\sim$ 2.2 galaxies will grow until they are `mass quenched' with increasing probability as they grow in mass beyond the Schechter mass of $\sim$10$^{10.7-11}$ \msun \citep{ConWec09,Pen+10}, which requires a tripling of mass based on the median stellar mass of the galaxies in our sample. Given an average specific star formation rate of $\sim$ 2 Gyr$^{-1}$, this process would take around 1.3 Gyr. Thus the most massive of our z $\sim$ 2.2 SFG sample will be effectively quenched by z $\sim$ 1.5. Indeed, highly unstable and morphologically disturbed SFGs are more rarely seen at this later epoch \citep{Kas+12}, also owing to the fact that galactic gas accretion has slowed by this time \citep{BirDekNei07}. The remainder may evolve into L$^{*}$ galaxies, as suggested by \cite{Con+08}.

Another possibility for the evolution of dispersion dominated objects is that they are the product of clump migration and coalescence at the centers of larger disks, and are therefore the descendants of rotation dominated galaxies. However, if the clumps are formed by gravitational instability, we would expect them to be continuously produced in the unstable, gas-rich disks, and we should see some brighter emission outside the center of the dispersion dominated galaxies. In addition, this scenario is at odds with the lower stellar masses, younger ages and lower metallicities of the dispersion dominated systems. Indeed, one would expect that the first galaxies to experience clump coalescence would be the most massive rather than the least massive ones.

\section{Conclusions}

This paper presents, for the first time, a side-by-side comparison of high quality deep AO and seeing-limited kinematic data of the same galaxies for a larger number of objects. We note that this unprecedented data leads us to similar conclusions as those made in previous work. Based on IFU data of 81 star forming galaxies at z=1--2.5, we compare the sub-samples of galaxies known as dispersion and rotation dominated. We find that the characterization of a galaxy into one of these kinematic groups is a strong function of the galaxy size. Small galaxies are much more likely to fall in the category of dispersion dominated galaxies due to insufficiently resolved rotation (especially with seeing-limited observations) and also as a result of the almost constant floor of velocity dispersion across all sizes paired with the linear increase of rotation velocity with size. Many galaxies that are considered dispersion dominated from more poorly resolved data actually show evidence for rotation with higher-resolution data. 

Despite the finding that galaxies characterized as dispersion-dominated often show evidence for rotation with higher spatial resolution data, they have different average properties than rotation dominated galaxies. They tend to have lower stellar and dynamical masses, higher gas fractions, younger ages and slightly lower metallicities.  We suggest that these galaxies could be precursors or `seeds' to larger rotating galaxies, as they accrete more mass onto the outer regions of their disks. Our AO-based results provide important insights for the analysis and interpretation of seeing-limited IFU data, such as will become available for large samples with KMOS, through the quantitative assessment presented of the effects of beam-smearing on the observed kinematics of real galaxies. \\

\acknowledgements
\small{
We thank the referee for a detailed and thoughtful review which led to substantial improvements in the paper. We thank the ESO staff, especially those at Paranal Observatory, for their ongoing support during the many past and continuing observing runs over which the SINS project is being carried out. We also acknowledge the SINFONI and PARSEC teams, whose devoted work on the instrument and laser paved the way for the success of the SINS observations. SFN is supported by an NSF grfp grant. CM and AR and GZ acknowledge partial support by the ASI grant ``COFIS-Analisi Dati'' and by the INAF grant ``PRIN-2008'' and ``PRIN-2010''. }


\nocite{Hop+12}
\nocite{Che+10}
\nocite{Gen+11}
\nocite{Kor+12}
\nocite{Tac+10}
\nocite{Erb+06}
\nocite{Dad+10}
\nocite{New+12a}
\nocite{Mur+05}
\nocite{OppDav06}
\nocite{OppDav08}
\nocite{HecArmMil90}
\nocite{Vei+05}
\nocite{OstShe11}
\nocite{Pet+00}
\nocite{Mar05}
\nocite{Wei+09}
\nocite{Ste+10}
\nocite{Dav+11}
\nocite{Cec+01}
\nocite{Vei+94}
\nocite{Wes+07}
\nocite{Law+12a}
\nocite{Law+12b}
\nocite{Sha+09}
\nocite{For+09}
\nocite{Man+11}
\nocite{Cal+00}
\nocite{BruCha03}
\nocite{Sch+04}
\nocite{Dav+07}
\nocite{Kom+11}
\nocite{Cha03}
\nocite{Eis+03}
\nocite{Bon+04}
\nocite{Shapley+03}
\nocite{Ken98}
\nocite{Noe+07}
\nocite{Dad+07}
\nocite{Rod+11}
\nocite{Pen+10}
\nocite{CowHuSon95}
\nocite{vdBer+96}
\nocite{ElmElmShe04}
\nocite{Elm+09}
\nocite{ElmElm05}
\nocite{ElmElm06}
\nocite{Gen+06}
\nocite{Gen+08}
\nocite{Law+07}
\nocite{Law+09}
\nocite{Law+12c}
\nocite{For+11a}
\nocite{Wuy+12}
\nocite{For+06}
\nocite{Wri+07}
\nocite{Wri+09}
\nocite{Sha+08}
\nocite{Cre+09}
\nocite{vSta+08}
\nocite{Epi+09}
\nocite{LemLam10}
\nocite{Jon+12}
\nocite{Wis+12}
\nocite{Dad+08}
\nocite{Tac+08}
\nocite{Swi+11}
\nocite{Nog99}
\nocite{Imm+04a}
\nocite{Imm+04b}
\nocite{BouElmElm07}
\nocite{ElmBouElm08a}
\nocite{DekSarCev09}
\nocite{Bou+10}
\nocite{Ker+05}
\nocite{Ker+09}
\nocite{DekBir06}
\nocite{Bow+06}
\nocite{KitWhi07}
\nocite{OcvPicTey08}
\nocite{Dave+08}
\nocite{DekSarCev09}
\nocite{Ose+10}
\nocite{Kom+11}
\nocite{Ste+04}
\nocite{Ade+04}
\nocite{Dad+04b}
\nocite{Kon+06}
\nocite{Lil+07}
\nocite{Abr+04}
\nocite{Cim+08}
\nocite{Kur+09}
\nocite{Ken98}
\nocite{Cal+00}
\nocite{Fan+88}
\nocite{Cal+94}
\nocite{Cal+96}
\nocite{MasKun99}
\nocite{May+04}
\nocite{Cid+05}
\nocite{Tac+06}
\nocite{Bou+07}
\nocite{Fel+12}
\nocite{Ken+07}
\nocite{Dad+10b}
\nocite{Gen+10}
\nocite{PetPag04}
\nocite{Wis+11}
\nocite{Epi+12}
\nocite{Pue+07}
\nocite{Pue+09}
\nocite{Nei+08}
\nocite{Wei+06}
\nocite{Kas+07}
\nocite{BouElmMar09}
\nocite{Age+09}
\nocite{Cev+10}
\nocite{Tre+04}
\nocite{Mai+08}
\nocite{Nel+12}
\nocite{MoMaoWhi98}
\nocite{Pen+10}
\nocite{For+13}
\nocite{Wuy+11a}
\nocite{Tac+12}
\nocite{BirDekNei07}
\nocite{Kas+12}
\nocite{New+12b}
\nocite{Man+09}
\nocite{Abu+06}
\nocite{Kur+13}
\nocite{vSta+08}
\nocite{Aum+10}
\nocite{GenDekCac12}
\nocite{Swi+12a}
\nocite{Swi+12b}
\nocite{Lop+12}
\nocite{ElmElm06}
\nocite{Cev+12}
\nocite{CacDekGen12}
\nocite{Lop+12}
\nocite{Cev+12}
\nocite{Bar+13}

\nocite{Pen+02}
\nocite{Nel+12b}
\nocite{KewEll08}
\nocite{Kur+09}
\nocite{Abr+04}
\nocite{Dad+04}
\nocite{Big+08}
\nocite{Tac+10}
\nocite{Dad+10}
\nocite{ConWec09}
\nocite{vdKruAll78}
\nocite{Con+08}
\nocite{Con+12}
\nocite{Ver+12}
\nocite{Que+12}
\nocite{Gon+10}
\nocite{Sai+12}
\nocite{KruBur10}
\nocite{ElmBur10}
\nocite{Gen+12}

\end{document}